\documentclass[twocolumn]{article}
\usepackage{icqproc,epsfig}

\def\(#1){(\ref{#1})}		

\def\kv{{\bf k}}
\def\qv{{\bf q}}
\def\bv{{\bf b}}

\def\rv{{\bf r}}
\def\dv{{\bf d}}

\def\g{\gamma}

\def\eps{\epsilon}
\def\G{\Gamma}
\def\sv{{\bf S}}
\def\GS{G_{\bf S}}
\def\GSk{G_{\bf S}^{\bf k}}
\def\={{\equiv}}

\def\cross{\!\times\!}

\begin{document}
\thispagestyle{empty}

\twocolumn[
\vspace*{30mm}
\begin{LARGE} 
\begin{center}
%
%
Magnetic quasicrystals: What can we expect to see in their neutron
diffraction data?
%
%
\end{center}
\end{LARGE}
\begin{large}
\begin{center} 
%
%
Ron Lifshitz
%
%
\end{center}
\end{large}
\begin{footnotesize}
\begin{it}
\begin{center}
%
%
Condensed Matter Physics 114-36, California Institute of Technology, Pasadena, CA 91125, U.S.A.
%
%
\end{center}
\end{it}
\end{footnotesize}
\begin{footnotesize}
\begin{center}
%
%
Received 1 September 1999
%
%
\end{center}
\end{footnotesize}
\vspace{4ex}
\begin{small}
\hrule\vspace{3ex}
\begin{minipage}{\textwidth}
{\bf Abstract}\vspace{2ex}\\
\hp 
%
%
The theory of magnetic symmetry in quasicrystals is used to
characterize the nature of magnetic peaks, expected in elastic neutron
diffraction experiments. It is established that there is no
symmetry-based argument which forbids the existence of quasiperiodic
long-range magnetic order. Suggestions are offered as to where one
should look for the simplest kinds of antiferromagnetic quasicrystals.
%
%
\vspace{2.5ex}
\end{minipage}\vspace{3ex}
\hrule
\end{small}\vspace{6ex}
]

%
%

\section{Introduction}

In recent years, we have witnessed a careful experimental
investigation of the question of long-range magnetic order in
rare-earth based icosahedral
quasicrystals~\cite{charier,islam,sato}. Nevertheless, discussions of
this matter have been somewhat unclear as to the actual nature of the
magnetic order one would expect to see in antiferromagnetic (AF)
quasicrystals, if they were to exist.  A partial answer to this
question can be obtained from a theory of the symmetry of magnetically
ordered quasicrystals~\cite{prl}. I intend to show here that such a
theory not only provides a valuable tool for analyzing neutron
diffraction data, but also helps to narrow down the possible magnetic
ordering one would expect to see in the classes of quasicrystals that
are known to exist today. I hope that this will help in guiding the
continuing search for new quasicrystals with this unique physical
property.

\section{The spin density field and its symmetry}

A magnetically-ordered crystal, whether periodic or aperiodic, is most
directly described by its spin density field $\sv(\rv)$. This field is
a 3-component real-valued function, transforming like an axial vector
under $O(3)$ and changing sign under time inversion.  One may think of
this function as defining a set of classical magnetic moments, or
spins, on the atomic sites of the material. For quasiperiodic crystals
the spin density field may be expressed as a Fourier sum with
a countable infinity of wave vectors
\begin{equation}\label{FourierS}
\sv(\rv)=\sum_{\kv\in L} \sv(\kv) e^{i\kv\cdot\rv} .
\end{equation}
The set $L$ of all integral linear combinations of the wave vectors in
\(FourierS) is called the {\it magnetic lattice.} Its {\it rank\/} $D$
is the smallest number of wave vectors needed to generate it by
integral linear combinations.  For quasiperiodic crystals, by
definition, the rank is finite. For the special case of periodic
crystals the rank is equal to the dimension $d$ of physical space.

In elastic neutron scattering experiments, every wave vector $\kv$ in
$L$ is a candidate for a magnetic Bragg peak whose intensity is given
by 
\begin{equation}\label{intensity}
I(\kv)\propto|\sv(\kv)|^2 - |\hat{\kv}\cdot\sv(\kv)|^2,
\end{equation} where $\kv$ is the scattering wave vector and
$\hat{\kv}$ is a unit vector in its direction. I have shown
elsewhere~\cite{krakow} that under generic circumstances there can be
only three reasons for not observing a magnetic Bragg peak at $\kv$
even though $\kv$ is in $L$: (a) The intensity $I(\kv)\neq 0$ but is
too weak to be detected in the actual experiment; (b) The intensity
$I(\kv)=0$ because $\sv(\kv)$ is parallel to $\kv$; and (c) The
intensity $I(\kv)=0$ because magnetic symmetry requires the Fourier
coefficient $\sv(\kv)$ to vanish. I shall explain below exactly how
this symmetry requirement, or ``selection rule,'' comes about.

The theory of magnetic symmetry in quasiperiodic crystals, which is
described in more detail in Ref.~\cite{prl}, is a reformulation of
Litvin and Opechowski's theory of spin space groups~\cite{litvin}.
Their theory, which is applicable to periodic crystals, is extended to
quasiperiodic crystals by following the ideas of Rokhsar, Wright, and
Mermin's ``Fourier-space approach'' to crystallography~\cite{rwm}. At
the heart of this approach is a redefinition of the concept of
point-group symmetry which enables one to treat quasicrystals directly
in physical space~\cite{volc}.  The key to this redefinition is the
observation that point-group rotations (proper or improper), when
applied to a quasiperiodic crystal, do not leave the crystal invariant
but rather take it into one that contains the same spatial
distributions of bounded structures of arbitrary size.

This generalized notion of symmetry, termed ``indistinguishability,'' is
captured by requiring that any symmetry operation of the magnetic
crystal leave invariant all spatially-averaged autocorrelation
functions of its spin density field $\sv(\rv)$ for any order $n$ and for
any choice of components $\alpha_i\in \{x,y,z\}$,
\begin{eqnarray}\label{corr2}
 \lefteqn{C^{(n)}_{\alpha_1\ldots\alpha_n}(\rv_1,\ldots,\rv_n)}
 \nonumber \\
 & & =\lim_{V\to\infty}{1\over V}\int_V d\rv 
   S_{\alpha_1}(\rv_1-\rv)\cdots S_{\alpha_n}(\rv_n-\rv).\qquad
\end{eqnarray} 

I have shown in the Appendix of Ref.~\cite{rmp} that an equivalent
statement for the indistinguishability of any two quasiperiodic spin
density fields, $\sv(\rv)$ and $\sv'(\rv)$, is that their Fourier
coefficients are related by
\begin{equation}\label{chidef} 
\sv'(\kv) = e^{2\pi i\chi(\kv)}\sv(\kv), 
\end{equation} 
where $\chi$, called a {\it gauge function,} is a real-valued scalar
function which is linear (modulo integers) on $L$. Only in the case of
periodic crystals can one replace $2\pi \chi(\kv)$ by $\kv\cdot\dv$,
reducing indistinguishability to the requirement that the two crystals
differ at most by a translation $\dv$.

With this in mind, we define the {\it point group $G$\/} of the
magnetic crystal to be the set of operations $g$ from $O(3)$ that
leave it indistinguishable to within rotations $\g$ in spin space,
possibly combined with time inversion. Accordingly, for every pair
$(g,\g)$ there exists a gauge function, $\Phi_g^\g(\kv)$, called a
{\it phase function}, which satisfies
\begin{equation}\label{phase}
\sv(g\kv) = e^{2\pi i \Phi_g^\g(\kv)}\g\sv(\kv).
\end{equation} Since $\sv([gh]\kv)=\sv(g[h\kv])$, one easily
establishes that the transformations $\g$ in spin space form a group
$\G$ and that the pairs $(g,\g)$ satisfying the point-group condition
\(phase) form a subgroup of $G\times\G$ which we call the {\it spin
point group $\GS$}. The corresponding phase functions, one for each
pair in $\GS$, must satisfy the {\it group compatibility condition,}
\begin{equation}\label{GCC}
\forall (g,\g), (h,\eta)\in\GS:\ \Phi_{gh}^{\g\eta}(\kv) \=
\Phi_g^\g(h\kv) + \Phi_h^\eta(\kv),
\end{equation} where ``$\=$'' denotes equality modulo integers. A
{\it spin space group,} describing the symmetry of a magnetic crystal,
whether periodic or aperiodic, is thus given by a magnetic lattice
$L$, a spin point group $\GS$, and a set of phase functions
$\Phi_g^\g(\kv)$, satisfying the group compatibility condition \(GCC).

\section{The diffraction pattern: A thinned-out magnetic lattice or
a shifted nuclear lattice?} 

I said earlier that every wave vector in the magnetic lattice is a
candidate for a diffraction peak unless symmetry forbids it. We are
now in a position to understand how this happens.  Given a wave vector
$\kv\in L$ we examine all spin point-group operations $(g,\g)$ for
which $g\kv=\kv$.  These elements form a subgroup of the spin point
group, called the {\it little spin group of $\kv$}, $\GSk$. For
elements $(g,\g)$ of $\GSk$, the point-group condition \(phase) can be
rewritten as
\begin{equation}\label{eigen}
\g\sv(\kv) = e^{-2\pi i \Phi_g^\g(\kv)}\sv(\kv).
\end{equation} 
This implies that the Fourier coefficient $\sv(\kv)$ is required to be
a simultaneous eigenvector of all spin transformations $\g$ in the
little spin group of $\kv$, with the eigenvalues given by the
corresponding phase functions. If a non-trivial 3-dimensional axial
vector satisfying Eq.~\(eigen) does not exist then $\sv(\kv)$ will
necessarily vanish. If such an eigen vector does exist its form might
still be constrained to lie in a particular subspace of spin space.

Of particular interest are spin transformations $\g$ that leave the
spin density field indistinguishable without requiring any rotation in
physical space. These transformations are paired in the spin point
group with the identity rotation $e$ and form a normal and abelian
subgroup of $\G$ called the {\it lattice spin group $\G_e$.}  In the
special case of periodic crystals, the elements of $\G_e$ are spin
transformations that when combined with translations leave the
magnetic crystal invariant.

The lattice spin group plays a key role in determining the outcome of
elastic neutron scattering, for if a magnetic crystal has a nontrivial
lattice spin group $\G_e$ then $\{e\}\cross\G_e \subseteq\GS^\kv$ for
every $\kv$ in the magnetic lattice, restricting the form of all the
$\sv(\kv)$'s. This may result in a substantial thinning-out of the
magnetic lattice, whereby only a fraction of the wave vectors give
rise to actual magnetic Bragg peaks.  Because this thinning of the
magnetic lattice is often quite extensive, it is common practice to
describe the magnetic peaks not as a thinned-out magnetic lattice but
rather in terms of the nuclear lattice $L_0$ (the one observed above
the magnetic ordering temperature) which is shifted by so-called
``magnetic propagation vectors.'' These two descriptions are in fact
equivalent and with some care can be used interchangeably.

\section{Where should we look?}

In the past I have tabulated all the decagonal spin space
groups~\cite{icq5}, as well as all the lattice spin groups for
icosahedral quasicrystals~\cite{prl}. In the latter case I also listed
explicitly, for every wave vector $\kv$ in the magnetic lattice,
whether through Eq.~\(eigen), symmetry requires $\sv(\kv)$ to vanish
or to take any special form. In a future publication I plan to provide
complete tables of spin space groups and the requirements which they
impose on neutron scattering experiments for all the relevant
quasiperiodic crystal systems (octagonal, decagonal, dodecagonal, and
icosahedral). 

Clearly, the theory of spin space groups provides a helpful tool for
analyzing neutron diffraction experiments. It lists the patterns of
magnetic Bragg peaks, compatible with each symmetry class, which can
then be directly compared with experiment. But on a more basic level
this theory answers one of the fundamental questions that have been
debated in recent years, which is whether it is even possible to have
long-range quasiperiodic magnetic order.  It establishes that even
though symmetry may impose constraints on the possible forms of
magnetic order one can have in a given quasicrystal, it clearly does
not forbid the existence of such order. {\it Thus, there is no
symmetry-based argument which disallows long-range magnetic order in
quasicrystals.}

Why is it then, that we have not yet observed unequivocal long-range
magnetic order in a quasicrystal? It might be because {\it
energetic\/} considerations lead to local frustration and spin-glass
ordering; It might be due to some other {\it physical\/} argument; Or
it might be simply because we have not found it yet. If this is the
case, then a more practical question to ask of a theory of magnetic
symmetry is whether it can offer any suggestions as to where to look
for such order.  Indeed, symmetry considerations may assist us in
deciding in which quasicrystal systems to look first for the {\it
simplest\/} kind of non-trivial magnetic ordering. Such ordering would
be the quasiperiodic analog of a simple AF periodic crystal where half
the spins are pointing ``up'' and the other half are pointing
``down.'' Symmetry arguments can guide us to those systems where such
ordering is possible.

I therefore close this essay with a short discussion of what this
quasiperiodic AF order looks like, followed by the list of systems
which are compatible with such order. It would then be up to the
metallurgists and material scientists to find the right chemical
systems which can sustain local magnetic moments and at the same time
are likely to have stable phases in these crystal systems.

\section{The quasiperiodic antiferromagnet}

The simple AF crystal, whether periodic or aperiodic, has a lattice
spin group $\G_e$ containing only two elements: the identity operation
$\eps$ and time inversion $\tau$. In the case of time inversion the
selection rule \(eigen) becomes
\begin{equation}\label{rule}
\tau\sv(\kv) = -\sv(\kv) = e^{-2\pi i \Phi_e^\tau(\kv)}\sv(\kv),
\end{equation}
which requires $\sv(\kv)$ to vanish unless $\phi_e^\tau(\kv)\=1/2$.
On the other hand, application of the group compatibility condition
\(GCC) to $(e,\tau)^2=(e,\eps)$ gives two possible values for this
phase,
\begin{equation}\label{values}
\phi_e^\tau(\kv)\=0 {\rm \ or\ } \frac12.
\end{equation}
It is not too difficult to show that exactly half of the wave vectors
in the magnetic lattice $L$ have $\phi_e^\tau(\kv)\=0$ and will
therefore not appear in the neutron diffraction pattern. These wave
vectors constitute a sublattice $L_0$ of index 2 in $L$. One can then
describe the set of wave vectors appearing in the diffraction diagram
either as the magnetic lattice $L$ without all the wave vectors in
$L_0$, or as $L_0$ shifted by $\qv$, where $\qv$, a ``magnetic
propagation vector,'' is any vector in $L$ which is not in the
sublattice $L_0$. In the simplest scenario $L_0$ is also the nuclear
lattice but this is not necessarily the case.

Consider a 1-dimensional spin chain with this lattice
spin group. If the chain is periodic then its (Fourier) magnetic
lattice is given by all integral multiples of a single wave vector
$\bv^*$ (I will keep the superscript-$*$ as a reminder that we are in
Fourier space). Because phase functions are linear it suffices to
specify the value of $\phi_e^\tau$ on $\bv^*$ and that will determine
its value on any wave vector in the lattice. Of the two possible
values \(values) the first, $\phi_e^\tau(\bv^*)\=0$ will result
through the selection rule \(rule) in $\sv(\kv)$ being zero everywhere
and therefore $\sv(\rv)=0$ as well. The only non-trivial assingment is
therefore $\phi_e^\tau(\bv^*)\=1/2$ which through the selection rule
\(rule) implies that all lattice wave vectors that are even multiples
of $\bv^*$ will be missing, or ``extinct,'' from the diffraction
pattern.

If the spin chain is quasiperiodic, say having a rank of 2, then its
magnetic lattice will be given by all integral linear combinations of
two wave vectors, $\bv^*_1$ and  $\bv^*_2$, whose magnitudes are
incommensurate. In this case the phase function $\phi_e^\tau$ is fully
determined by specifying its two independent values on $\bv^*_1$ and
$\bv^*_2$. At first glance it would seem as if there are three
distinct non-trivial assignments of values given by
\begin{equation}\label{rank2}
\bigl(\phi_e^\tau(\bv^*_1),\phi_e^\tau(\bv^*_2)\bigr)\ \=\
(0,\frac12) {\rm \ or\ } (\frac12,0) {\rm \ or\ } (\frac12,\frac12).
\end{equation}
It turns out that these three assignments are equivalent, leading to
the same spin space group, due to the fact that for a quasiperiodic
chain one has the added freedom of changing the basis of the
magnetic lattice. A basis transformation from $(\bv^*_1, \bv^*_2)$ to
$(\bv^*_1+\bv^*_2, \bv^*_2)$ or to $(\bv^*_1, \bv^*_1+\bv^*_2)$ takes
one, respectively, from the first or second assignment in \(rank2) to
the third. Thus, the diffraction pattern of a quasiperiodic AF spin
chain can always be described as a magnetic lattice given by wave
vectors of the form $\kv = n_1\bv^*_1 + n_2\bv^*_2$ where all vectors
with $n_1+n_2$ {\it even\/} are extinct. Equivalently, it may
described as a lattice $L_0$, generated by the wave vectors
$\bv^*_1+\bv^*_2$ and $\bv^*_1-\bv^*_2$, and shifted by the vector
$\bv^*_1$.

\begin{figure}[t]
\begin{center}
\epsfig{file=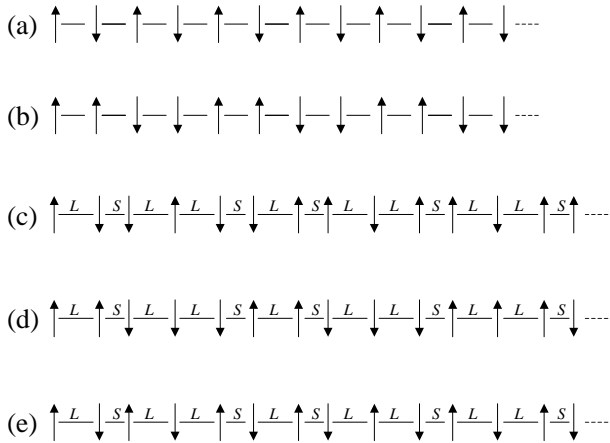,width=\columnwidth}
\end{center}
\caption{Examples of antiferromagnetic spin chains. (a) and (b) are
periodic with different magnetic unit cells. (c)-(e) are the AF
Fibonacci chains, obtained as described in the text by using a
modified grid method. In (c) spins separated by a long ($L$) segment
are anti-parallel and those separated by a short ($S$) segment are
parallel; in (d) the opposite occurs; and in (e) all nearest-neighbor
pairs are anti-parallel.}
\label{fig1} 
\end{figure}

Knowing the different possibilities in Fourier space allows us to
immediately construct simple direct-space examples of AF spin chains
having these symmetries.  Figures~\ref{fig1}(a) and (b) show two
periodic AF spin chains in which the ``magnetic unit cell'' is twice
or four-times as large as the ``nuclear unit cell.''  Both of these
chains will exhibit the same magnetic diffraction peaks, the only way
to distinguish them being a direct comparison with the nuclear
diffraction pattern, which can be obtained above the magnetic ordering
temperature. Figures~\ref{fig1}(c)-(e) show three AF Fibonacci chains,
obtained by setting the ratio $b^*_1/b^*_2$ to the golden mean
$(1+\sqrt5)/2$, and using the three different assignment of the phase
function values given in \(rank2). Again, as discussed above, all
three are expected to have the same magnetic diffraction peaks and the
only way to distinguish them is a comparison with the nuclear
diffraction pattern.

Which of the actual quasicrystal systems that are known to exist today
allow simple AF order? Axial quasicrystals admit two kinds of simple
AF order.  Since they are all quasiperiodic in the plane normal to the
$n$-fold axis and periodic along this axis it is always possible to
have periodic AF order along the $n$-fold axis. This would give an AF
quasicrystal but not in the true sense that we are interested in. Only
when $n$ is a power of 2 is it possible to have true quasiperiodic AF
order in the plane normal to the $n$-fold axis~\cite{rmp}. {\it Thus,
among the known axial quasicrystals one should concentrate the search
for simple AF order in the octagonal crystal system.}

Only two of the three Bravais classes in the icosahedral system admit
simple AF order~\cite{prl}. Such order is possible if the nuclear
lattice is either simple (giving a magnetic lattice which is
body-centered in Fourier space) or if the nuclear lattice is
face-centered in Fourier space (giving a simple icosahedral magnetic
lattice). Unfortunately, most of the known icosahedral quasicrystals,
including the rare-earth based ones, are face-centered in direct space
and therefore do not allow simple AF order. Furthermore, icosahedral
quasicrystals which are body-centered in direct space are not yet
known to exist. {\it Thus, in the icosahedral system, one should look
for simple AF order in crystals that have a simple icosahedral nuclear
lattice.}

%
%

\hp

\begin{footnotesize}
\begin{frenchspacing}

\end{frenchspacing}
\end{footnotesize}

\end{document}